\documentstyle[aps,prl,epsfig]{revtex}
\begin{document}
\draft
\def\etal{{\em et\ al.\/}, }
\def\eg{{\em e.g.\/}, }
\def\ie{{\em i.e.\/}, }
\def\beq{\begin{equation}}
\def\eeq{\end{equation}}
\def\beqarr{\begin{eqnarray}}
\def\eeqarr{\end{eqnarray}}
\def\ix{\int_{-L/2}^{L/2}dx\,}
\def\ixi{\int dx\,}
\def\ii{\int_{-\infty}^{\infty}}
\def\eps{\varepsilon}
\def\ov{\over}
\def\ol{\overline}
\def\sgn{{\rm sgn}}
\def\ph{\hat{\phi}}
\def\Pt{\tilde{\Psi}}
\def\tr{{\rm tr}}
\def\det{{\rm det}}
\def\la{\langle}
\def\ra{\rangle}
\def\8{\infty}
\def\one{{\bf 1}}
\def\lp{\lambda'}
\def\Zint{
  {\mathchoice
    {{\sf Z\hskip-0.45em{}Z}}
    {{\sf Z\hskip-0.44em{}Z}}
    {{\sf Z\hskip-0.34em{}Z}}
    {{\sf Z\hskip-0.35em{}Z}}
    }
  }
\def\del{\partial}
\def\delx{\partial_x}
\def\delxp{\partial_{x'}}
\def\delt{\partial_{\tau}}
\def\delT{\partial_t}
\def\XLL{{$\chi$LL}}
\def\gtlt{\,{\stackrel {\scriptscriptstyle >}{\scriptscriptstyle <}}\,}
\def\bh{{\hat \beta}}
\def\ph{{\hat \phi}}
\def\XSG{{$\chi$SG }}
\def\bJ{{\bf J}}
\def\Lt{L^{(1/2)}}
\def\Lh{L'^{(1/2)}}
\def\lh{{\hat \lambda}}
\def\nb{{\bar \nu}}
\def\tb{{\bar \tau}}
\def\ID{I}
\def\Ab#1{{\bf A}^{(#1)}}
\def\Bb#1{{\bf B}^{(#1)}}
\def\Cb#1{{\bf C}^{(#1)}}
\def\Af#1{\tilde{\bf A}^{(#1)}}
\def\Bf{\tilde{\bf B}}
\def\Cf#1{\tilde{\bf C}^{(#1)}}
\def\Rf#1{\tilde{\bf R}^{(#1)}}
\def\Gf#1{\tilde{\bf G}^{(#1)}}
\def\X#1{2\pi(x-v_{#1}t+i\epsilon_t)}
\def\XP#1{2\pi(x-x'-v_{#1}t+i\epsilon_t)}
\def\Tt{{\tilde T}}
\def\Lt{{\tilde L}}
\def\mat{\sf}
\def\Wt{{\widetilde W}}
\twocolumn[\hsize\textwidth\columnwidth\hsize\csname
@twocolumnfalse\endcsname
\title{The fractional quantum Hall effect in infinite layer systems}

\author{J. D. Naud$^1$, Leonid P. Pryadko$^2$ and S. L. Sondhi$^1$.}

\address{
$^1$Department of Physics,
Princeton University,
Princeton, NJ 08544, USA
}

\address{
$^2$School of Natural Sciences,
Institute for Advanced Study,
Olden Lane,
Princeton, NJ 08540, USA}

\date{\today}
\maketitle
\begin{abstract}
Stacked two dimensional electron systems in transverse
magnetic fields exhibit three dimensional fractional quantum Hall
phases. We analyze the simplest such phases and find novel
bulk properties, e.g., irrational braiding. These phases
host ``one and a half'' dimensional surface phases in which motion
in one direction is chiral. We offer a general analysis of conduction in
the latter by combining sum rule and renormalization group arguments,
and find that when interlayer tunneling is marginal or irrelevant they
are chiral semi-metals that conduct only at $T > 0$ or with disorder.
\end{abstract}
\pacs{}
]
Low dimensional electron systems exhibit striking examples of strongly
correlated behavior.  In the search for higher dimensional analogs, a
tempting strategy is to couple such systems weakly in the hope of
achieving a ``dimensional continuation'' of the strong correlation
physics \cite{dimcont}. In this letter we report our results on
several aspects of the continuation of the fractional quantum Hall
effect to a three dimensional setting.  The central idea behind such a
continuation is this: in the absence of interlayer tunneling, an
infinite stack of two dimensional electron systems can exhibit
``multi-component'' quantum Hall behavior which generalizes the
possibility already known to be realized in bilayers or systems where
spin is important \cite{gm1}. Such behavior will be accompanied by a
gap which will then allow a weak tunneling to be turned on without
destroying continuity. This route to three dimensional quantum Hall
phases comes with an added bonus, namely that the chiral edge states
existing in each layer will hybridize and yield a family of ``one
and a half'' dimensional phases that will live on the surfaces of the
three dimensional systems and exhibit interesting transport in the
direction transverse to the layers.

In the context of the integer effect, this possibility was first
explored experimentally \cite{stormer1} and has more recently been
studied systematically by both theory \cite{chiraltheory} and
experiment \cite{chiralexpt} leading not only to a demonstration of
three dimensional quantum Hall behavior but also of the formation of a
chiral metal at its surface via hybridization of the (Fermi liquid)
edge states of the individual layers. The extension of fractional QH
behavior has attracted less attention.  Balents and Fisher
\cite{chiraltheory} commented on the edge dynamics of uncorrelated
fractional layers but more central to many of our concerns is the very
early work of Qiu, Joynt, and MacDonald (QJM)\cite{QJM} on the
possibility of {\it interlayer correlated} states and the evaluation
of their energetic stability.

We begin by describing the states that we study, and note some unusual
features of the bulk physics such as non-trivial quasiparticle
structure, irrationally distributed charge and a corresponding
statistics. Next we formulate the edge theory for these states and
show that in the clean limit a very general sum rule argument can be
used to exactly determine the conductivity along the stack (the
$z$-axis), $\sigma^{zz}$.  This argument implies metallic
behavior at finite temperature and disorder in cases where the
tunneling is relevant and semi-metallic behavior in cases where it is
marginal or irrelevant. This connection relies on a rigid
correspondence between RG relevance and the form of the ground state
which is specific to chiral systems. We illustrate the general
argument by computations of the conductivity at weak tunneling in
which the disorder is treated exactly. In much of the analysis we will
be particularly interested in the QJM ``131'' state (defined below)
which is perhaps the simplest correlated infinite layer state.
Technical details of most assertions and related material on other
chiral many-body systems will follow in a separate publication
\cite{nps4}.

\noindent
{\bf States:}
Consider a system which consists of $N$ parallel layers of 2DEGs in a
strong perpendicular magnetic field that we assume freezes out any
interesting spin dynamics as well. We will be interested in the
generalized Laughlin-Halperin states, 
\beqarr
\lefteqn{\Psi_{0}(\{z_{i,\alpha}\})=\prod_{i=1}^N\prod_{\alpha < \beta}^{N_i}
(z_{i\alpha}-z_{i\beta})^{K_{ii}}}&&\nonumber\\
&&{}\times
\prod_{i<j}^N\prod_{\alpha=1}^{N_i}\prod_{\beta=1}^{N_j}
(z_{i\alpha}-z_{j\beta})^{K_{ij}}
e^{-\sum_{\alpha,i}|z_{\alpha,i}|^2/4}
\label{eq:multilaugh}.
\eeqarr
Here $z_{i\alpha}$ is the coordinate of electron $\alpha$ in layer
$i$, and $N_i$ is the number of electrons in layer $i$.
The exponents are specified by a symmetric $N\times N$ matrix $K$,
which we will take to be tridiagonal $K_{ij}=
m\delta_{ij}+n(\delta_{i,j-1}+\delta_{i,j+1})$ in the interests of
simplicity and plausibility. Clearly, the diagonal elements determine 
the intralayer and the off-diagonal elements specify the interlayer 
correlations. In the large $N$ limit of interest, it is convenient
to assume periodic boundary conditions in the $z-$direction 
\cite{fn-caveat} whereupon the filling factor in each layer of
this ``$nmn$'' state is $\nu=1/(m+2n)$. Evidently, more than one
state can occur at the same filling. The competition between the
states 050 and 131 at filling $\nu=1/5$ per layer is especially
interesting as the simplest example of a potential transition between
interlayer uncorrelated and correlated states. We note that 
QJM found that 050 gives way to 131 as the interlayer separation
was decreased. In what follows we will impose one further restriction
in order to obtain the simplest edge dynamics---we will require that
the states give rise only to co-propagating edge modes, which
is equivalent to requiring that $K$ posses only positive eigenvalues
or that $n<m/2$.

\noindent
{\bf Quasiparticles:} 
By analogy to the single-layer quantum Hall effect we can construct a
state with a single quasihole at point $\xi$ in layer $j$ as:
\beq
\label{eq:qhole}
\Psi_{(j,\xi)}(\{z_{i\alpha}\})=\prod_{\alpha=1}^{N_j}(z_{j\alpha}-\xi)
\Psi_0(\{z_{i\alpha}\}).
\eeq
A standard plasma screening argument yields a total charge of
$Q=\sum_k Q^{(j)}_k=\sum_k (K^{-1})_{jk}=\nu_j=1/(m+2n)$ for the
quasihole, where $Q^{(j)}_k$ is the charge in layer $k$ due to the
quasihole in layer $j$.
  The distribution of this charge, first noted by QJM, is
rather more interesting:
\beq
\label{eq:indcharges}
Q^{(j)}_k={1\ov\sqrt{m^2-4n^2}} 
\left({\sqrt{m^2-4n^2}-m\ov 2n}\right)^{|k-j|}.
\eeq
For interlayer correlated states ($n > 0$), the individual charges 
in the layers are {\em irrational}. This result, for
an isolated quasihole, is stable to the inclusion of weak tunneling.

A closer look at the plasma screening computation yields an
interesting structure for the quasihole in that the charge in layers
farther from the nominal location of the quasihole, though smaller, is
spread over larger areas.  For example, for the 131 state, the mean
squared radius goes as $\langle r^2_j\rangle = {4\,\left( 3 +
    {\sqrt{5}}\,j \right) }/{5} $ for a quasihole centered in layer
$0$. Finally, we also readily obtain the braiding phase for a
quasihole in layer $k$ as it encircles the planar position of a
quasihole in layer $j$, \beq
\label{eq:berry2}
\gamma_B^{({\rm stat})} = -2\pi Q^{(j)}_k,
\eeq
which is again irrational. 

\begin{figure}[htbp]
  \begin{center}
    \leavevmode
    \epsfxsize=0.7\columnwidth
    \epsfbox{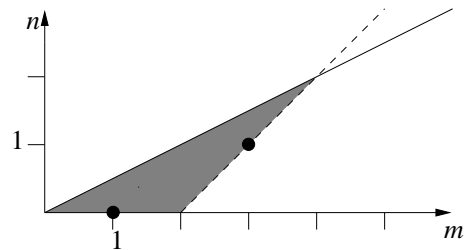}\vskip1mm
    \caption{The $m$-$n$ plane.  States below the solid line are maximally
      chiral.  The dashed line separates states with irrelevant
      tunneling (below) from states with relevant tunneling (above).
      The shaded region contains all maximally-chiral states with
      non-irrelevant tunneling: 010 and 131.}
    \label{fig:relstab}
  \end{center}
\end{figure}

\noindent
{\bf Edge Theory:}
The edge theory of the $N$-layer quantum Hall state contains $N$
chiral bosons $u_i(x)$ whose commutation relations are determined by
the same matrix $K$ which specifies the correlation exponents in the
bulk wavefunction\cite{wen}
\beq
\label{eq:uCR}
[u_i(x), u_j(x')]=i\pi K_{ij}\,{\rm sgn}(x-x').
\eeq
In the absence of tunneling the low-energy effective Hamiltonian of
the edge theory is
\beq
\label{eq:H0u}
{\cal H}_0=\ix {1\over{4\pi}}V_{ij} 
\,:\!\delx u_i\,\delx u_j\!:,
\eeq
where $L$ is the length of the edge and $V$ is a symmetric, positive
definite, $N\times N$ matrix which depends on the interactions and
confining potentials at the edge.  The electron annihilation operator
at the edge of layer $i$ is $\Psi_i(x)\propto e^{iu_i(x)}$, and the
Hamiltonian density for tunneling between layers $i$ and $i+1$ is
$\lambda\Psi^{\dag}_i(x)\Psi_{i+1}(x)+\,{\rm h.c.}$, where $\lambda$ is
the tunneling amplitude which we take to be uniform along the edge.
The lowest-order perturbative RG flow of the tunneling amplitude for
the $nmn$ state is
\beq
\label{eq:RGflow}
{d\lambda\ov d\ell}=\left({2-m+n}\right)\lambda.
\eeq
Combining this result with the condition for maximum chirality,
$n<m/2$, we obtain the diagram in Fig.~\ref{fig:relstab}.  There are
only two maximally chiral multilayer states for which interlayer
electron tunneling is not irrelevant: 010 with relevant tunneling and
131 with marginal tunneling. (Also of interest are the bosonic 020
state with marginal tunneling and the states 121 and 242 which have
positive semi-definite $K$ matrices and hence a non-trivial mixing of
edge and bulk excitations \cite{nps4}.)

\noindent
{\bf Sum Rule and $\mathbf \sigma^{zz}$:}
As we mentioned above, the case of relevant tunneling, 010, has been
previously studied, so we will concentrate on the case of marginal
tunneling, 131.  First consider the multilayer edge theory in the
absence of disorder.  By evaluating the double commutator of the full
Hamiltonian (including tunneling) with the $z$-axis Fourier transform of
the charge density operator $\rho_i(x)=(K^{-1})_{ij}\delx u_j(x)/2\pi$
integrated over all $x$, we can derive the following exact sum rule
\beq
\label{eq:sumrule}
\int d\omega \, \Re e\, \sigma^{zz}(\omega)=
-{\pi d\ov NL}\la {\cal H}_{\lambda}\ra,
\eeq
where $d$ is the interlayer separation and ${\cal H}_{\lambda}$ is
the tunneling part of the Hamiltonian.

In the absence of tunneling, the ground state of the Hamiltonian
${\cal H}_0$ is known, and the expectation value on the
r.h.s.\ of the sum rule is zero.  The marginality of the
tunneling perturbation for the 131 state implies that there exists a
finite range of tunneling amplitudes $\lambda$ for which the ground
state is stable, which can be understood as follows.  The
tunneling Hamiltonian ${\cal H}_{\lambda}$ commutes with ${\cal H}_0$,
and therefore first-order degenerate perturbation theory is exact, and
the eigenvalues of ${\cal H}_0+{\cal H}_{\lambda}$ are linear
functions of $\lambda$.  The case of marginal tunneling corresponds to
a dimensionless $\lambda$, and hence the energy of the first excited
state behaves like $E_1(\lambda)- E_1(0) \sim \lambda/L$.  If
$E_0(\lambda)$ is the ground state energy, then $E_1(0)-E_0(0)$
must be $\sim 1/L$ since the edge theory is gapless, and we can conclude
that the first level crossing occurs at $\lambda\sim 1$.  Therefore,
for a finite range of $\lambda$ the ground state for $\lambda\neq 0$
is identical to the ground state for $\lambda=0$.  In contrast, for
the case of relevant tunneling, $\lambda$ has the dimensions of an
energy and an analogous line of reasoning shows that the first level
crossing occurs at a value of $\lambda$ that approaches zero as the
system size $L$ is taken to infinity, indicating the ground state is
not stable.  Note that these arguments are non-perturbative in the
tunneling amplitude.  This reasoning can be made precise,
and the region of stability in $\lambda$ bounded\cite{nps4}.

Combining the stability of the ground state, the sum rule, and the
general condition that $\Re e\, \sigma^{zz}(\omega)$ is positive
semi-definite, we can conclude that at $T=0$: $\sigma^{zz}(\omega)=0$
for all $\omega$ \cite{fn-effecth}.  In the absence of disorder the
surface of the 131 state exhibits insulating behavior in the direction
perpendicular to the layers.  The presence of either a finite
temperature or disorder would make the expectation value on the
r.h.s.\ of the sum rule non-zero, and therefore lead to a
non-vanishing conductivity.  This property, in particular that adding
disorder {\em increases} the conductivity, is reminiscent of a
semi-metal and hence the surface of the 131 state may be considered a
``chiral semi-metal.''

\begin{figure}[htbp]
  \begin{center}
    \leavevmode
    \epsfxsize=0.85\columnwidth
    \epsfbox{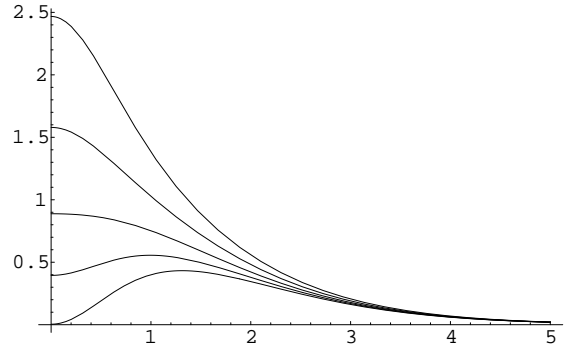}\vskip1mm
    \caption{
    The real part of the vertical conductivity for the 131 multilayer
    with disorder (\protect\ref{eq:sigmazz/w/dirt}).  The horizontal
    axis is frequency measured in units of the disorder parameter
    $\Delta$.  The temperatures of the curves, starting with the
    uppermost, are: $T/\Delta=0.25,0.2,0.15,0.1,0.01$.  }
    \label{fig:sigmazz} \end{center}
\end{figure}

\noindent
{\bf $\mathbf \sigma^{zz}$ with disorder:} To illustrate the above
claim, consider adding disorder to the multilayer edge theory in the
form of random scalar potentials $V_i(x)$ which couple to the edge
charge density $\rho_i(x)$ in each layer.  We assume $V_i(x)$ is a
Gaussian random variable uncorrelated between different layers, \ie
$\ol{V_i(x)V_j(x')}=\delta_{ij}Z(x-x')$, where the overbar denotes
disorder averaging.  The conductivity is evaluated via a Kubo formula.
Since the Hamiltonian with tunneling is a theory of $N$ interacting
chiral bosons, the calculation is performed perturbatively in
$\lambda$, but the disorder is treated exactly.  We expect that for
the disordered case the lowest-order result in $\lambda$ is
reliable\cite{nps4}.  To illustrate our results it is especially
convenient to use a specific short-ranged disorder potential
correlator $Z(x)$ characterized by a single energy $\Delta$ which
determines the strength of the disorder and the correlation length of
the random potential along the edges $\ell_0\sim 1/\Delta$. It is also
useful to choose the matrix $V$ appearing in ${\cal H}_0$ proportional
to $K^{-1}$.  Physically, this corresponds to a specific non-zero
value of $g$, the nearest-neighbor density-density coupling between
the layers. With these choices we find, 
\beq
\label{eq:sigmazz/w/dirt}
\Re e \,\ol{\sigma^{zz}}(\omega)={\lambda^2 d\ov 48\pi^2}
{\omega^2+4\pi^2T^2\ov \Delta\cosh\left({\pi \omega/2\Delta}\right)}.
\eeq
This result is shown in Fig.~\ref{fig:sigmazz} at various
temperatures.  At zero temperature the conductivity vanishes in the DC
limit.  At low temperatures the maximum occurs at a finite frequency,
while at higher temperatures the conductivity is peaked around
$\omega=0$.  It can be shown that the large and small frequency
asymptotics of the above result for $\ol{\sigma^{zz}}(\omega)$ are not
significantly modified by perturbing the matrix $V$ away from this
point\cite{nps4}.

\noindent
{\bf Clean limit:} 
The 131 state in the clean ($\Delta=0$)
 limit is characterized by
two dimensionless parameters, $\lambda$ and $g$, and  is therefore quantum
critical.  Its finite temperature conductivity is thus of particular
interest as there are very few cases where the transport in such a
regime is known\cite{damle-sachdev}.  If we take the limit of a clean
system $\Delta\rightarrow 0$ in Eq.~(\ref{eq:sigmazz/w/dirt}) we find:
\beq
\label{eq:cleansigmazz}
\Re e\, \sigma^{zz}(\omega)\propto\lambda^2 d\,T^2\delta(\omega).
\eeq 
This is consistent with our result that $\sigma^{zz}$ in the clean
system vanishes at $T=0$.  An interesting aspect of this form is that
it is proportional to $\delta(\omega)$, \ie the conductivity vanishes
at all $\omega\neq 0$, and the DC conductivity is infinite.  One
question to ask is whether these features are an artifact of keeping
only the first non-vanishing term in a perturbative expansion in
$\lambda$.  Although we cannot definitively answer the question about
the finiteness of the DC conductivity beyond ${\cal O}(\lambda^2)$, we
have proven that at higher orders in $\lambda$ the conductivity cannot
vanish at all $\omega\neq 0$.  The proof proceeds by showing that
$\sigma^{zz}(\omega)$ is zero for all $\omega\neq 0$ if and only if
the current operator $I^z$ commutes with the Hamiltonian, and then
demonstrating that for the 131 state this commutator is
non-vanishing\cite{nps4}.  On general grounds we expect that the
$\delta(\omega)$ factor in Eq.~(\ref{eq:cleansigmazz}) is broadened by
both higher-order corrections in the tunneling $\lambda$, and
nearest-neighbor density-density interactions $g$, but the exact form
remains to be found.

\noindent
{\bf Irrelevant Cases:}
The perturbative calculation of the $z$-axis conductivity in the
presence of tunneling and disorder can be extended to states with
irrelevant tunneling, \ie $nmn$ states with $m>n+2$.  These states
exhibit similar behavior to the marginal case of the 131 state, in
particular, in the absence of disorder they are perfectly insulating
in the $z$-direction at $T=0$.  In the presence of disorder we find
$\ol{\sigma^{zz}}\sim \omega^{\alpha}$ as $\omega\rightarrow 0$ at
zero temperature and $\ol{\sigma^{zz}}\sim T^{\alpha}$ as
$T\rightarrow 0$ at zero frequency, where the exponent
$\alpha=2(m-n-1)$.  For the case of states without interlayer
correlations, $n=0$, this temperature scaling of the DC conductivity
is different from that found by Balents and
Fisher\cite{chiraltheory} who considered the case where
interactions rather than disorder provide a mechanism for scattering.
Essentially, this difference in the temperature scaling arises because
in the clean case the scattering is determined by a dimensionless
interaction strength $g$, while in the disordered case the scattering
depends on the dimensionful parameter $\Delta$.  

\noindent
{\bf 050 versus 131:}
Note that in the presence of disorder the $z$-axis DC conductivity
scales as $T^8$ for the 050 state, while for the 131 state it scales
as $T^2$.  These states occur at the same electron density, $\nu=1/5$
per layer.  In their numerical work, QJM found a phase transition
between these states as the layer separation $d$ was varied\cite{QJM}.
The significant difference in the temperature scaling of the $z$-axis
DC conductivity between 050 and 131 suggests a way to experimentally
detect this phase transition.

\noindent
{\bf Spectrum:} 
In our analysis of the surface conduction we have been fortunate
 to make progress without detailed knowledge of the excitation
spectrum. The latter would clearly be desirable in the cases where
tunneling is not irrelevant. We digress to remark that in the simplest
non-trivial case of $N=2$ layers, the edge theory in the presence of
tunneling and disorder {\it can} be solved exactly via
fermionization\cite{nps1-2}.  In the bilayer case the interesting
experimental probe is not the $z$-axis conductivity but rather the
collective mode structure, \eg the density response function between
layers $i$ and $j$: ${\cal D}^R_{ij}(t,x)=
-i\theta(t)\la[\rho_i(t,x),\rho_j(0)]\ra$.  For the
bilayer analog of the multilayer 131 state, conventionally referred to
as 331, we find that ${\cal D}^R$ contains two signals which propagate
at distinct velocities\cite{nps1-2}.  This is of experimental interest
because, in contrast, the bilayer Pfaffian state, which occurs at the
same filling $\nu_{\rm tot}=1/2$ as the 331 state, yields only a
single signal in ${\cal D}^R$.  Therefore, a time resolved measurement
of the density response function at the bilayer edge using two
spatially separated electrodes could conclusively determine whether
the experimentally observed Hall plateau at $\nu_{\rm tot}=1/2$ is the
331 state or the Pfaffian.

The solution of the infinite layer 131 edge problem and its cousins 
is not as easily found.  However, we have succeeded in fermionizing
this family of problems---generically, by adding extra free bosonic
degrees of freedom. For example, for the 131 state, the fermionized
Hamiltonian is:
\begin{eqnarray}
    {\cal H}_{131}&=&\ix\biggl\{ \sum_{j,\alpha}\tilde{V}_{jj}
    :\!\psi_{j+\alpha}^\dagger i\delx
    \psi_{j+\alpha}\!:\nonumber\\
    &&{}+{1\over2}\sum_{\alpha,i\neq j}\tilde V_{ij}
    (:\!\psi_{i+\alpha}^\dagger \psi_{i+\alpha}\!:)
    \,(:\!\psi_{j+\alpha}^\dagger 
    \psi_{j+\alpha}\!:)
    \nonumber\\
    & &{}+\sum_j\left(\lambda\, \psi_{j-1/2}^\dagger \psi_{j}^\dagger
    \,\psi_{j+1} 
    \psi_{j+3/2}+{\rm h.c.}\right)\biggr\}, 
  \label{eq:fermionized-131}  
\end{eqnarray}
where $\psi_j$ are $2N$ species of chiral fermions with half-integer
indices, the matrix $\tilde{V}\equiv KV$, and $\alpha=0,1/2$.  We
anticipate that these fermionic formulations may serve as a point of
departure for constructing exact solutions. We note that these are
``one and a half'' dimensional problems, and hence their solution
would appear to be (half) a step up from the solution of strictly one
dimensional problems.
             
\noindent{\bf Acknowledgments:}
We would like to acknowledge support by an NSF Graduate Research
Fellowship (JDN), DOE Grant DE-FG02-90ER40542 (LPP) as well as NSF
grant No.\ DMR-99-78074, US-Israel BSF grant No.\ 9600294, and
fellowships from the A.~P.~Sloan Foundation and the David and Lucille
Packard Foundation (SLS).  LPP and SLS thank the Aspen Center for
Physics for its hospitality during the completion of part of this
work.


\begin{thebibliography}{99}
  
\bibitem{dimcont} Examples of such work include D.~L.~Cox and A.~
  Zawadowski, Adv. Phys., {\bf 47} 599 (1998); P. W. Anderson, {\em
    The theory of superconductivity in the high-Tc cuprates\/}
  Princeton Univ.\ Press, Princeton (1997); V.~J.~Emery \etal {\tt
    cond-mat/0001077}.
  
\bibitem{gm1} S.~M. Girvin and A.~H. Mac{D}onald, in {\em Novel
    Quantum Liquids in Low-Dimensional Semiconductor Structures}, ed.\ 
  S.~DasSarma and A.~Pinczuk (Wiley, New York, 1995).

\bibitem{stormer1}
H. L. Stormer \etal Phys.\ Rev.\ Lett.\ {\bf 56}, 85 (1986). 

\bibitem{chiraltheory}
J. T. Chalker and A. Dohmen, Phys.\ Rev.\ Lett.\ {\bf 75}, 4496 (1995);
L. Balents and M. P. A. Fisher, Phys.\ Rev.\ Lett.\ {\bf 76}, 2782
(1996); J. J. Betouras and J. T. Chalker, {\tt cond-mat/0005151} and
references therein.

\bibitem{chiralexpt}
D. P. Druist \etal
%%, P. J. Turley, K. D. Maranowski, E. G. Gwinn, and
%%A. C. Gossard, 
Phys.\ Rev.\ Lett.\ {\bf 80}, 365 (1998).

\bibitem{QJM}
X. Qiu, R. Joynt, and A. H. MacDonald, Phys.\ Rev.\ B {\bf 40}, 11943
(1989); Phys.\ Rev.\ B {\bf 42}, 1339 (1990).

\bibitem{nps4} J. D. Naud, Leonid P. Pryadko and S. L. Sondhi, unpublished,
  {\tt cond-mat/0006506}.

\bibitem{fn-caveat}
In finite stacks with open boundary conditions there are
non-uniformities in $\nu$ near the end layers, see Ref.~\cite{nps4}.

\bibitem{wen}
X.-G. Wen and A. Zee, Phys.\ Rev.\ B {\bf 46}, 2290 (1992).

\bibitem{fn-effecth}
This is a statement about the surface of the system within the region of
validity of the low-energy effective Hamiltonian (\ref{eq:H0u}).

\bibitem{damle-sachdev}
K. Damle and S. Sachdev, Phys.\ Rev.\ B {\bf 56}, 8714 (1997).

\bibitem{nps1-2} J. D. Naud, L. P. Pryadko, and S. L. Sondhi, Nucl.\ 
  Phys.\ B {\bf 565}, 572 (2000); unpublished,
  {\tt cond-mat/0006173}.

\end{thebibliography}
\end{document}